\documentclass[reprint,
superscriptaddress,
onecolumn,
 amsmath,amssymb,
 aps,
pra,
]{revtex4-2}
\usepackage[T1]{fontenc}
\usepackage{url}
\usepackage{graphicx}
\usepackage{epstopdf}
\usepackage{epsfig}
\usepackage{xcolor}
\usepackage{subcaption}
\usepackage{dcolumn}
\usepackage{bm}
\usepackage{bbold}

\usepackage{caption}
\usepackage{mathrsfs}
\usepackage{amsmath,amssymb}

\usepackage[left=1.5cm, right=1.5cm, top=1.785cm, bottom=2.0cm]{geometry}
\usepackage{graphicx} 
\usepackage{xcolor}
\usepackage{url}
\usepackage{indentfirst}
\usepackage{hyperref,xcolor}

\let\oldeq\eqref 
\renewcommand\eqref{Eq. \oldeq}

\providecommand{\DIFdel}[1]{} 

\hypersetup{
    colorlinks=true,
    linkcolor=cyan,
    filecolor=magenta,
    citecolor=cyan,
    urlcolor=cyan,
    pdfpagemode=FullScreen,
    }

\begin{document}

\preprint{APS/123-QED}

\title{Magnetically Coupled Circuits to Capture Dynamics of Ionic Transport in Nanopores}

\author{Filipe Henrique}
\affiliation{Department of Chemical and Biological Engineering, University of Colorado, Boulder}
\author{Ankur Gupta}
\email{Corresponding author:
ankur.gupta@colorado.edu}
\affiliation{Department of Chemical and Biological Engineering, University of Colorado, Boulder}
\date{\today}

\begin{abstract}
Ionic transport within charged nanopores is commonly represented by resistor-capacitor transmission line circuits, where charging electrical double layers are modeled as capacitors, and the resistance to ionic current is modeled as resistors. However, these circuits fail to account for oscillations observed in experimental Nyquist plots of impedance, which are attributed ad hoc to effects such as complex porous structures or chemical reactions. Here, we show that diffusivity asymmetry between ions in confinement -- overlooked in previous studies -- produces Nyquist plots with two turns. Additionally, we demonstrate that ionic transport is more accurately described by magnetically coupled inductor-resistor circuits than by a simple resistor-capacitor circuit. Our results show that an impedance response of ionic transport in nanopores for arbitrary Debye lengths is better captured by two Warburg elements in parallel than a single Warburg element. 
\end{abstract}

\maketitle

\preprint{APS/123-QED}

\section{Introduction}

The physics of ionic transport in nanoporous electrodes underpins electrochemical technologies such as supercapacitors \cite{simon2020perspectives,wang2021recent}, capacitive deionization \cite{porada2013review} and fuel cells \cite{tang2014tailored}. A key feature of ionic transport is the formation of electrical double layers (EDLs) inside charged pores \cite{lasia2014electrochemical,wang2021recent}, which have been described using effective circuits for more than 7 decades, beginning from the seminal work of de Levie~\cite{de1963porous,de1964porous}. The approach of effective circuits is powerful because it enables one to distill the physical processes through distinct circuit elements, which subsequently helps characterize and optimize the design of electrodes and electrolytes. 
\par{} The equivalent circuits are typically made up of a ladder with rungs of resistors and capacitors, also known as a transmission line (TL). The capacitors, with capacitance per unit length $C$, represent the formation of an EDL next to the pore surface, whereas the resistors, with resistance per unit length $R$, represent the resistance to ionic currents through the pore. For a small applied potential and thin double layers~\cite{huang2020editors,wu2022understanding}, TLs yield a Warburg open pore impedance~\cite{pedersen2023equivalent} $\mathcal{W}_o=R\coth(\sqrt{\omega RC})/\sqrt{\omega RC}$, or a Nyquist plot with a single visible curvature maximum between the 90$^\circ$ low-frequency and 45$^\circ$ high-frequency asymptotes. This apparent curvature change occurs at a characteristic frequency $\omega\sim 1/\left(RC \right)$, corresponding to EDL formation over the electrode length \cite{orazem2017electrochemical}. However, experiments frequently report more complex behaviors that are not captured by a single Warburg element. These dependencies are attributed ad hoc to complex porous structures and chemical reactions, and the underlying physics remains an open question.  
\par{} For instance, previous works have measured the impedance response of activated carbon nanofiber structures, fitted using a modified TL circuit \cite{real2022analyses}, and polyacrylonitrile supercapacitor electrodes \cite{nabil2020preparation}. In both studies, changes in the sign of the impedance curvature with frequency were observed for intermediate frequencies in Nyquist plots \citetext{\citealp[Fig.~8]{real2022analyses}; \citealp[Fig.~12]{nabil2020preparation}}. In the former, they were attributed to anomalous dispersion, but additional physical processes have not been investigated.

\par{} Several modifications of de Levie's TL have been developed to model a finite pore length \cite{posey1966theory,janssen2021transmission}, variations in pore shape~\cite{de1965influence,keiser1976abschatzung,black2010pore}, overlapping double layers~\cite{gupta2020charging, henrique2022charging}, pore networks~\cite{song1999electrochemical,itagaki2007impedance,itagaki2010complex,biesheuvel2010nonlinear,lian2020blessing,henrique2024network}, surface conduction~\cite{mirzadeh2014enhanced}, and electrochemical reactions~\cite{de1964porous,biesheuvel2011diffuse}, among others. While the elements of the circuit may change, a single-TL structure is typically conserved.
The robustness of this representation raises the question: is a TL for the electric potential always sufficient to describe the dynamics of ionic transport at low potentials? In this paper, we show that it is not; in fact, representing the ionic transport of asymmetric electrolytes in nanopores requires coupled circuits. We achieve this representation through magnetically coupled circuits that reduce to a single classical TL in the appropriate limits of symmetric electrolytes or thin double layers. Furthermore, the impedance response of the coupled circuits shows curvature oscillations that are qualitatively similar to those seen in experiments.

Electrolyte asymmetry, i.e., the inequality of ionic mobilities, is a usual and important feature of most electrolytes. For instance, some common  aqueous electrolytes employed in supercapacitors due to their high dissociation constants \cite{pal2019electrolyte}  are characterized by ionic diffusivity ratios, e.g., $D_+/D_-\approx 0.4$ for KOH and $\approx 8.7$ for H$_2$SO$_4$ \cite{vanysek1993ionic}. Unequal ionic mobilities give rise to a wealth of intriguing electrokinetic phenomena, such as colloidal levitation near electrode surfaces \cite{woehl2015bifurcation,hashemi2020perturbation,balu2021thin} and long-range repulsion between oppositely charged surfaces separated by an electrolyte \cite{richter2020ions}. Another important feature of ion transport in nanoporous electrodes is confinement, frequently manifested in high-capacitance carbon electrodes through pore sizes on the order of nanometers \cite{chmiola2006anomalous,zhang2014highly,yin2022effects}, which are comparable to the EDL thickness. We previously showed that the simultaneity of electrolyte asymmetry and overlapping double layers results in local enhancement or depletion of the total number of ions in a nanopore, even at low potentials, affecting charging times \cite{henrique2022impact}. de Levie's TL and subsequent modifications do not capture this effect and its impact on the charging dynamics. We address it in this paper.

\section{Physical Motivation}

\begin{figure}[ht!]
\includegraphics[width=\textwidth]{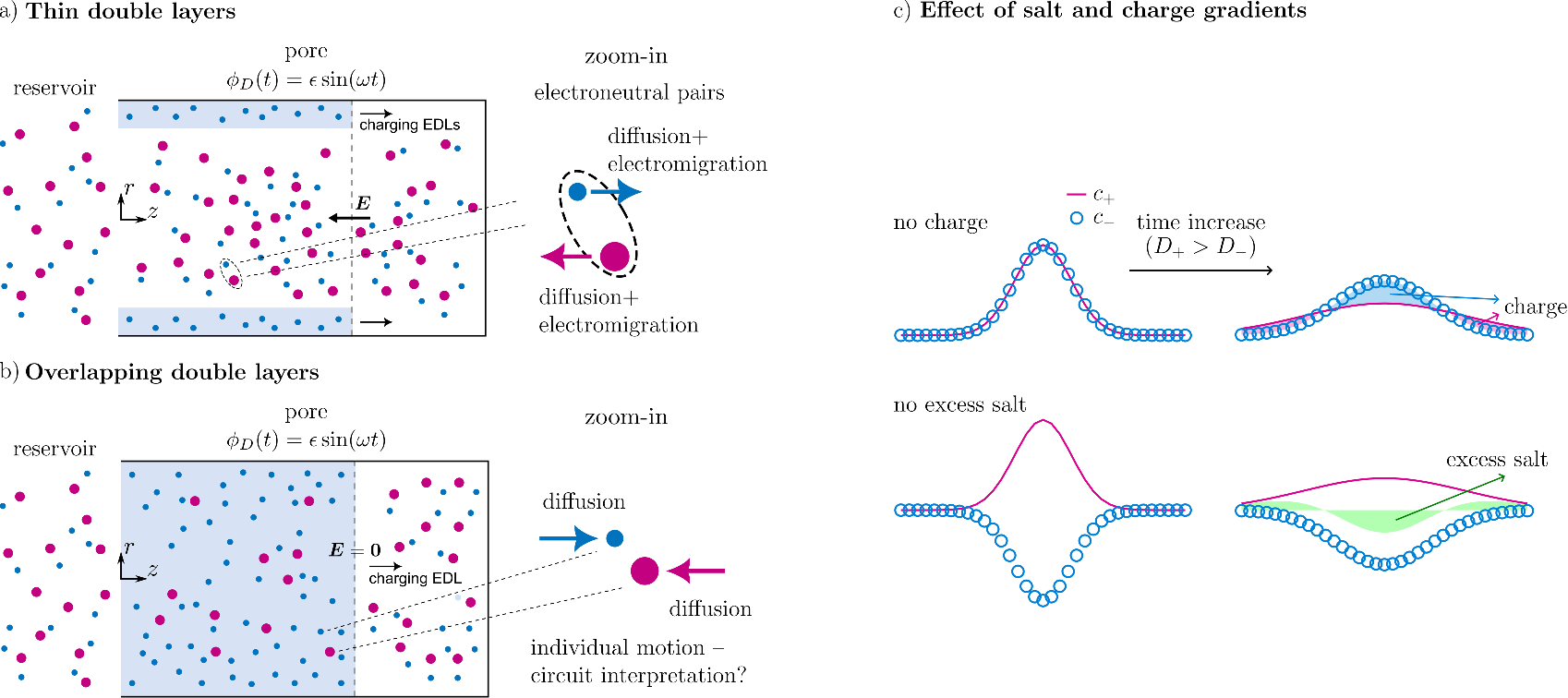}
\caption{\textbf{Effects of confinement on EDL charging of asymmetric electrolytes under an AC electrode potential} $\phi_D(t)=\epsilon\sin(\omega t)$. Cations in pink, anions in blue, EDLs in lighter blue. a) Thin EDLs: electric potentials are fully screened, so electromigration transports ions in pairs in the electroneutral bulk. b) Overlapping double layers: electroneutrality breaks down and ion transport is uncoupled. c) Schematic of diffusion of charge and salt gradients: mean-field densities of cations in pink lines and anions in blue orbs for an initial gradient of charge or salt. A mobility mismatch produces charge from salt gradients and salt from charge gradients.}
\label{fig:schematic}
\end{figure}

Before delving into the mathematical details, we physically motivate the need for coupled circuits and explain their particular importance for ions with asymmetric mobilities in the overlapping EDL regime. First, we focus on the case of a thin EDL. Figure \ref{fig:schematic}a illustrates this scenario for a binary electrolyte, with cations in pink and anions in blue. It represents a snapshot where the EDLs, in light blue, have formed partially along the symmetry axis of the pore. Due to the thinness of the EDLs, an uncharged bulk region exists; see, e.g., \cite{biesheuvel2010nonlinear}. Since electroneutrality must hold in the bulk, ions move in countercurrents as pairs, thus their transport is coupled.  Electromigration and diffusion are balanced in the bulk to enforce electroneutrality, thus the charge current can be expressed as a multiple of the electric field, i.e., as an Ohmic element. As the ions are transported in pairs, there is no dynamics of salt  (total ion concentration). Therefore, de Levie's TL remains applicable for asymmetric mobilities and thin EDLs. An exception occurs for large potentials, where surface conduction also drives transport \cite{chu2007surface,mirzadeh2014enhanced}. We will focus only on the linear limit due to its applicability to electrochemical impedance spectroscopy (EIS).

\par{} Figure \ref{fig:schematic}b illustrates the limit of overlapping EDLs. Here, the surface potential is only partially screened since the EDLs do not have room to develop fully. In other words, there is no bulk region. This feature relaxes the requirement of electroneutrality throughout the pore volume, so the ions do not need to move as pairs. As we show mathematically in Sec.~\ref{sec:math}, diffusion dominates electromigration, indicating that the charging process occurs at the diffusive timescales of both ions, $\ell^2/D_\pm$, where $\ell$ is the pore length. Effectively, the cation and anion fluxes become decoupled and driven by their respective mobilities. The ion concentrations are then only constrained by the double-layer structure. Finally, we are ultimately interested in the charge flux, but it is coupled to the salt flux \cite{henrique2022impact}. As illustrated schematically in Fig. \ref{fig:schematic}c, when electroneutrality breaks down and the ions are allowed to diffuse according to their mobilities, salt gradients produce charge and charge gradients produce salt. This implies that we need two circuits: one for charge and one for salt, and these circuits need to interact with each other. Next, we will derive these circuits from reduced-order governing equations of ion transport.  

\section{Mathematical Formulation}
\label{sec:math}

\subsection{Linearized Poisson-Nernst-Planck Equations}

\par{} We consider a cylindrical nanopore of radius $a$ filled with a binary electrolyte under a sinusoidal electrode potential; see Fig.~\ref{fig:schematic}a,b. The nanopore is in contact with an electroneutral reservoir and the ions dynamically move in and out of the pore in response to the evolving electrode charge. Assuming the linear limit of low applied potentials for the dead-end pore, we follow the common practice of neglecting convective transport; see, e.g., \cite{bazant2004diffuse,biesheuvel2010nonlinear,yaroshchuk2021interaction,ratschow2024convection} for limitations of the approach. To develop a model of the process, let us first define the shorthands $\bar{D}=(D_++D_-)/2$ for the average diffusivity, $\bar{z}=(|z_+|+|z_-|)/2$ for the average valence, and $I_{\infty}=|z_+|^2c_{+,\infty}+|z_-|^2c_{-,\infty}$ for the ionic strength of the reservoir, where the concentrations of the ions in the fully dissociated electrolyte are $c_{\pm,\infty}$. With these definitions, we scale the ion concentrations $c_{\pm}$ in the pore by $I_{\infty}/(2\bar{z}^2)$, the electric potential $\phi$ by the thermal voltage $\phi_{\mathrm{th}}=k_{B}T/(\bar{z}e)$, the gradient operator $\nabla$ by the inverse of pore length $1/\ell$, and time $t$ by $\ell^2/\bar{D}$. Based on these scales, we choose dimensionless parameters describing the pore geometry and electrolyte properties, namely, the diffusivity contrast $\beta=(D_+-D_-)/(D_++D_-)$, valence contrast $\gamma=(|z_+|-|z_-|)/(|z_+|+|z_-|)$, pore aspect ratio $a/\ell$, and pore radius relative to  Debye length, $\kappa = a/\lambda$, where $\lambda=\sqrt{\varepsilon k_{B}T/(e^2I_\infty)}$ and $\varepsilon$ is the electrolyte permittivity. Thus, the dimensionless Nernst-Planck (NP) equation is \cite{deen2012analysis,henrique2022impact}
\begin{equation}
    \frac{\partial c_\pm}{\partial t}=(1\pm\beta)\left[\nabla^2c_\pm\pm(1\pm\gamma)\nabla\cdot(c_\pm\nabla\phi)\right].
    \label{eq:pnp_c_pm}
\end{equation}
In the context of EIS, we analyze the equation for low potentials, where the ion concentrations in the electromigration term can be approximated by the reservoir concentrations $1/(1\pm\gamma)$ \cite{henrique2022impact}, yielding 
\begin{subequations}
\begin{equation}
    \frac{\partial c_\pm}{\partial t}=(1\pm\beta)\left(\nabla^2c_\pm\pm\nabla^2\phi\right).
    \label{eq:np_c_pm} 
\end{equation}
\eqref{eq:np_c_pm} is supplemented by Poisson's equation for a linear dielectric,
\begin{equation}
        -\nabla^2\phi =\left(\dfrac{\ell}{\lambda}\right)^2\dfrac{(1+\gamma)c_+-(1-\gamma)c_-}{2},
    \label{eq:lin_poisson_c_pm}
\end{equation}
such that \eqref{eq:pnp_c_pm} are the linearized Poisson-Nernst-Planck (PNP) equations.
\label{eq:pnp_c_pm}
\end{subequations}

We seek to determine the charge dynamics, so writing linear combinations of these equations for charge and salt density is useful. To this end, we define dimensionless densities of charge $\rho$ and \textit{excess} salt $c$ (relative to the total ion concentration of the reservoir), respectively scaled by $\varepsilon\phi_{\mathrm{th}}/\lambda^2$ and $I_{\infty}/\bar{z}^2$. In terms of the dimensionless ion concentrations, they are given by
\begin{subequations}
\begin{equation}
    \rho=\dfrac{(1+\gamma)c_+-(1-\gamma)c_-}{2}
\end{equation}
and
\begin{equation}
    c=\dfrac{c_++c_-}{2}-\dfrac{1}{1-\gamma^2}.
\end{equation}
It will also be important for the dimensionality reduction (averaging) of the governing equations in the pore to write the right-hand sides of the linearized NP \eqref{eq:np_c_pm} in terms of conserved quantities over a cross-section, such as the excess salt density. In fact, a linearized Boltzmann distribution over a cross-section yields a constant salt density, as the product of charge and electric potential is of the order of the applied potential squared \cite{henrique2022impact}.

Now, due to EDL formation, neither the charge density nor the electric potential is conserved over a cross-section. Thus, we will represent them in terms of an invariant property, constructed from the ionic electrochemical potentials in the dilute solution scaled by $2k_BT$,
\begin{equation}
    \mu_\pm=\dfrac{\ln c_\pm\pm (1\pm\gamma)\phi}{2}.
\end{equation}
We combine them into the excess electrochemical potential of charge \cite{henrique2024network}, constructed as a difference between the dimensionless electrochemical potentials of the ions and their electric potentials at the electrode-electrolyte interface. Linearizing for low potentials, the electrochemical potential of charge takes the form
\begin{equation}
    \mu=\mu_+-\mu_--\phi_D(t)\sim \rho+\phi-\phi_D(t),
    \label{eq:mu_def}
\end{equation}
\end{subequations}
up to a constant. Hence the linearized PNP equations for charge and salt densities are
\begin{subequations}
\begin{align}
    &\dfrac{\partial \rho}{\partial t}=(1+\beta\gamma)\nabla^2\mu+\beta(1-\gamma^2)\nabla^2c,\\
    &\dfrac{\partial c}{\partial t}=\beta\nabla^2\mu+(1-\beta\gamma)\nabla^2c,\\
    &-\nabla^2\phi=\left(\dfrac{\ell}{\lambda}\right)^2\rho.
    \label{eq: poisson_asym}
\end{align}
\label{eq:pnp_asym_rho_c}
\end{subequations}

In the main text, we neglect the reservoir resistance. Thus, the pore domain $V_p=\{\boldsymbol{x}=(r,\theta,z): 0\le r\le 1,\,0\le z\le 1\}$ is connected at $z=0$ to 
the electroneutral reservoir, which is unaffected by the applied potential, maintaining a reference potential $\phi(z=0^-)=0$ and the concentrations of ions resulting from dissociation $c_\pm(z=0^-)=1/(1\pm\gamma)$. The electrolyte in the pore charges radially and axially in response to the dimensionless applied potential $\phi_D=\epsilon\sin(\omega t)$ at $r=1$. Reservoir resistance is discussed through the inclusion of a static diffusion layer (SDL) in the Appendix.

\subsection{Averaged Equations}

\par{} We briefly extend the procedure of Refs. \cite{gupta2020charging,henrique2022charging,henrique2022impact,henrique2024network} to time-dependent applied potentials to derive reduced-order governing equations in the pore domain. Framing the current discussion in terms of invariant properties over a cross-section is crucial to the circuit representation that we will develop, which is the original contribution of the work. 

Assuming slender pores $a/\ell \ll 1$, we can invoke radial equilibrium \cite{yaroshchuk2011transport,gupta2020charging,aslyamov2022analytical} for frequencies of the applied potential below $O(a^2/D)$. Therefore, in the absence of Faradaic reactions, the radial fluxes of ions vanish over long times compared to axial diffusion \cite{yaroshchuk2011transport}. This can be written as the conservation of electrochemical potentials along any cross-section. This leads to the relations $\mu(r,z,t)=\mu(z,t)$ and $c(r,z,t)=c(z,t)$. Using these relations to average the PNP equations, we obtain \cite{henrique2022impact,henrique2024network}
\begin{subequations}
    \begin{align}
    &\dfrac{\partial\bar{\rho}}{\partial t}=(1+\beta\gamma)\dfrac{\partial^2\mu}{\partial z^2}+\beta(1-\gamma^2)\dfrac{\partial^2c}{\partial z^2},\\
    &\dfrac{\partial c}{\partial t}=\beta\dfrac{\partial^2\mu}{\partial z^2}+(1-\beta\gamma)\dfrac{\partial^2c}{\partial z^2},
    \end{align}
    \label{eq:pnpZ}
\end{subequations}
where a bar denotes a cross-sectional average, $\bar{f}=2\int_0^1 fr\,\mathrm{d}r$. By relating the average charge density to the electrochemical potential of charge, we can close the system of equations. To this end, we write a simplified form of Poisson's equation for long pores by asymptotically neglecting axial derivatives,
\begin{equation}
    -\dfrac{1}{r}\dfrac{\partial}{\partial r}\left(r\dfrac{\partial\phi}{\partial r}\right)=\kappa^2\rho,
    \label{eq:poissonR}
\end{equation}
where $\kappa=a/\lambda$ is the relative pore size, the dimensionless inverse Debye length. Using \eqref{eq:mu_def} for the radially invariant electrochemical potential of charge, we write the left-hand side of \eqref{eq:poissonR} in terms of charge density and solve it, finding
\begin{equation}
    \rho(r,z,t)=\mathcal{D}(\kappa)\bar{\rho}(z,t)\dfrac{I_0(\kappa r)}{I_0(\kappa)},
    \label{eq:rho_r_asym}
\end{equation}
where $I_n$ is the modified Bessel function of the first kind of order $n$. $\mathcal{D}(\kappa)=\kappa I_0(\kappa)/(2I_1(\kappa))$ can be interpreted as an effective diffusivity of charge for a symmetric electrolyte \cite{henrique2022charging,henrique2024network}. Evaluating \eqref{eq:mu_def} at the pore surface, where $\phi=\phi_D(t)$, and using \eqref{eq:rho_r_asym}, we find the relation $\mu(z,t)=\mathcal{D}\bar{\rho}$, whence the averaged PNP equations can be written as
\begin{subequations}
\label{eq:avg_pnp}
\begin{align}
&\frac{1}{\mathcal{D}(\kappa)}\frac{\partial \mu}{\partial t} =(1+\beta\gamma)\frac{\partial^2\mu}{\partial z^2}+\beta(1-\gamma^2)\frac{\partial^2c}{\partial z^2},
\label{eq:avg_pnp_mu} \\
&\frac{\partial c}{\partial t} =\beta\frac{\partial^2\mu}{\partial z^2}+(1-\beta\gamma)\frac{\partial^2c}{\partial z^2}.
\label{eq:avg_pnp_c}
\end{align}
\end{subequations} 
\eqref{eq:avg_pnp} are statements of conservation of charge and salt densities, respectively. Their interpretation as an equivalent circuit representation of asymmetric electrolytes in confinement will be the main subject of this paper. The equations underscore that the dynamical response of ionic transport in a nanopore can be expressed naturally by excesses of the electrochemical potential of charge and salt density. Both $\mu$ and $c$ only vary in the axial direction, but can be used to recover radial profiles through \eqref{eq:mu_def} and \eqref{eq:rho_r_asym}. For symmetric diffusivities, i.e., $\beta=0$, \eqref{eq:avg_pnp} decouple and the charge flux is only driven by derivatives of $\mu$, which encompass both electromigrative and diffusive fluxes
\cite{henrique2024network}.
\par{} For asymmetric diffusivities, $\beta \neq 0$, the charge and salt dynamics are coupled. For thin EDLs, i.e., $\kappa \gg 1$, the effective diffusivity is dominated by electromigration and large compared to molecular diffusion, $\mathcal{D} \gg 1$, setting a charge dynamics that is much faster than the salt dynamics. This effectively creates a weak coupling between the two equations with an impedance response given by a single Warburg element; this is discussed along with Fig.~\ref{fig:nyquist}. However, for $\kappa \lesssim O(1)$ and moderate dimensionless frequencies, the two equations are strongly coupled, $\mathcal{D}(\kappa) \sim 1$ due to molecular diffusion, such that the timescales of the charge and salt dynamics are both diffusive and thus comparable.

\begin{figure}[ht!]
\centering
\includegraphics[width=.8\textwidth]{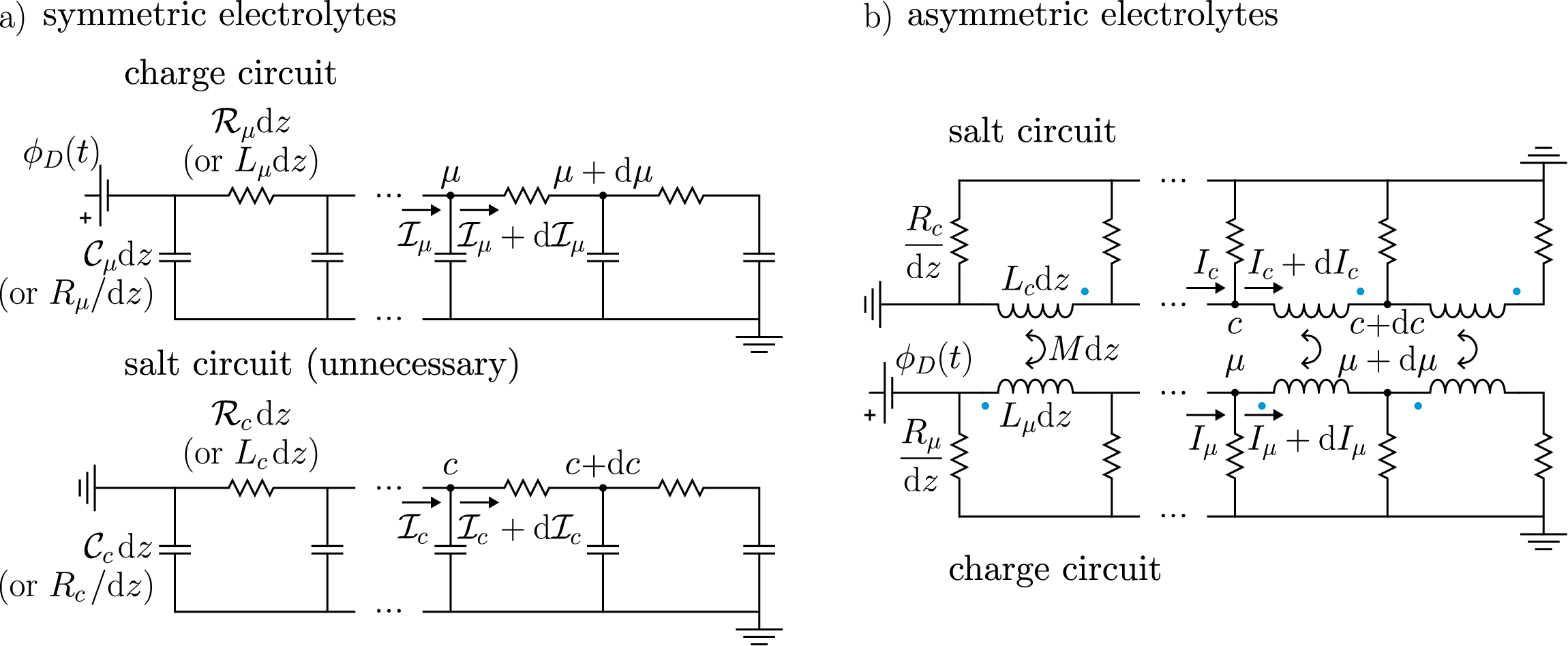}
\caption{\textbf{Circuit representation of asymmetric electrolytes in charged confinement}. a) A generalized dimensionless de Levie circuit in terms of the electrochemical potential of charge in a symmetric electrolyte \cite{henrique2024network}. b) Dimensionless magnetically coupled TL circuits for simultaneous charge and salt dynamics in a confined asymmetric electrolyte.}
\label{fig:circuits}
\end{figure}

\section{Circuit Interpretation}

\par{} Given the foundational importance of circuit representations to EIS experiments, we now derive one for \eqref{eq:avg_pnp} by resorting to well-established analogies between differential equations and equivalent circuits of infinitely many nodes \cite{macneal1949solution,pedersen2023equivalent,janssen2021transmission}, where the short distance between nodes allows one to approximate differences by differentials. To explain the rationale of the derivation, we first highlight an important feature of the known case $\beta=0$. There, \eqref{eq:avg_pnp} reduce to uncoupled unsteady transient diffusion equations for charge and salt transport. Thus, charge and salt can both be represented by independent generalized de Levie circuits, as seen in Fig. \ref{fig:circuits}a. However, since salt does not affect charge transport, it suffices to describe the latter. Before we move on, we note one distinction between Fig. \ref{fig:circuits}a and de Levie's original circuit: the voltage is the electrochemical potential of charge, $\mu$, not the electric potential $\phi$~\cite{henrique2024network}. This change makes the circuit applicable beyond thin EDLs, to an arbitrary dimensionless pore size $\kappa$. 

In the general case of asymmetric electrolytes, $\beta\ne 0$, we need to include the coupling terms on the right-hand side of the \eqref{eq:avg_pnp}, absent in de Levie's argument. Physically, these terms are charge fluxes produced by salt gradients and salt fluxes produced by charge gradients, as illustrated in Fig. \ref{fig:schematic}c. In a generalized de Levie circuit, they could be understood as the current in each circuit affecting the voltage differences in the other one. Nevertheless, we do not know of any electromagnetic mechanisms that promote such a coupling. Therefore, we cannot propose a direct extension of de Levie's circuit by simply adding these voltage drops to the resistors of Fig. \ref{fig:circuits}a due to currents in their correspondents in the other circuit. However, if instead of resistors, these circuits had inductors with mutual inductances between corresponding elements as in Fig. \ref{fig:circuits}b, there would be a coupling of voltage differences across an inductor due to a \textit{time derivative} of current in its neighbor per Lenz-Faraday's law. Thus, the key to capturing coupled charge and salt fluxes due to ion concentration gradients in the equivalent circuit will be substituting elements by corresponding ones with impedances multiplied by $i\omega$, e.g., having an inductor instead of a resistor, and associating the time derivatives of currents in the charge and salt circuits with the actual physical currents of charge and salt in the electrolyte.

Let us further illustrate the substitution of circuit elements with an example. Ohm's law gives $V_R=RI_R$ for a current $I_R$ and a voltage drop $V_R$ across a resistor of resistance $R$. If we had an inductor of inductance $L=R$ \footnote{In dimensionless variables.} instead of the resistor, voltage and current would be related by $V_L=RdI_L/dt$. With phasor representations of the sinusoidal AC variables denoted by hats, e.g., a current $I=\mathrm{Im}(\hat{I}\exp(i\omega t))$, the phasor of the time derivative of current would be $i\omega\hat{I}$ \cite{alexander2013fundamentals}. Thus, the impedance of the resistor is $\mathcal{Z}_R=\hat{V}_R/\hat{I}_R=R$ and the impedance of this inductor is $\mathcal{Z}_L=\hat{V}_L/\hat{I}_L=i\omega L=i\omega R=i\omega \mathcal{Z}_R$. Consequently, the impedance of the resistor can be inferred from the impedance of the inductor by $\mathcal{Z}_R=\mathcal{Z}_L/(i\omega)$. Effectively, that is the same as inferring the impedance of the resistor from the time derivative of current through the inductor, $\mathcal{Z}_R=\hat{V}_L/\widehat{(dI_L/dt)}$. That is, for the calculation of the impedance of the resistor, the time derivative of the current through the equivalent inductor plays the same role as the current through the resistor. Equivalently, we will use circuit elements with $i\omega$-multiplied impedances to determine the impedance of the original physical system. To our knowledge, the charge-salt flux coupling makes a circuit with modified impedances necessary, as we know no physical mechanism for coupling the resistors of the original de Levie circuit.

To apply the idea illustrated in the example to our circuit representation, we first multiply the frequency of all de Levie circuit elements by $i\omega$ and then allow for a mutual inductance between the correspondingly positioned inductors of both circuits; see Fig. \ref{fig:circuits}b. Therefore, \eqref{eq:avg_pnp} are understood as statements of time derivatives of Ohm's laws for the resistors of each circuit. With these changes, the impedance of the circuit will be related to that of the electrolyte by $\mathcal{Z}_{\mathrm{electrolyte}}=i\omega\mathcal{Z}_\mathrm{circuit}$. Magnetically coupled circuits are well-known in electrical engineering applications. Such a coupling may be undesirable, as in crosstalk of printed circuits, or desirable when the goal is transferring power from one circuit to another, as in directional couplers \cite{orfanidis2002electromagnetic}. These circuits have also been used for a long time to represent the vibration of bending beams \cite{macneal1949solution}. In the present context, magnetically coupled circuits such as the one we proposed may be considered in EIS analyses of electrochemical processes with coupled exponential relaxations.

\par{} We now show mathematically that the circuits of Fig. \ref{fig:circuits} are indeed equivalent to \eqref{eq:avg_pnp} for a specific choice of resistances and inductances.
Consider the LR circuits described in the previous paragraph and illustrated in Fig. 
\ref{fig:circuits}b. Applying Faraday's law to the inductors in the circuits of electrochemical potential and salt, respectively, where the dot convention \cite{alexander2013fundamentals} was used to indicate the orientation of the winding of the coils and corresponding sign of differences of potentials due to the mutual inductances, 
\begin{subequations}
\begin{eqnarray}
-\dfrac{\partial\mu}{\partial z}=L_\mu\dfrac{\partial I_\mu}{\partial t}-M\dfrac{\partial I_c}{\partial t},  
\label{eq:faraday_mu} \\
-\dfrac{\partial c}{\partial z}=-M\dfrac{\partial I_\mu}{\partial t}+L_c\dfrac{\partial I_c}{\partial t}.
\label{eq:faraday_c}
\end{eqnarray}
\eqref{eq:faraday_mu} and \eqref{eq:faraday_c} can be combined to write the time derivatives of the currents in terms of the gradients of both voltages, e.g.,
\begin{equation}
    \dfrac{\partial I_\mu}{\partial t}=-\dfrac{1}{L_\mu(1-k^2)}\left(\dfrac{\partial \mu}{\partial z}+\dfrac{M}{L_c}\dfrac{\partial c}{\partial z}\right),
    \label{eq:di_phi/dt}
\end{equation}
where $k=M/\sqrt{L_cL_\mu}$ is the magnetic coupling coefficient. Applying the time derivative of Ohm's law to the resistors of the electrochemical potential circuit,
\begin{equation}
\dfrac{\partial \mu}{\partial t}=-R_\mu\dfrac{\partial^2I_\mu}{\partial t\partial z},
\end{equation}
\label{eq:omhs+faraday}
\end{subequations}
and substituting into \eqref{eq:di_phi/dt}, we obtain
\begin{subequations}    
\begin{equation}
\dfrac{1}{R_\mu}\dfrac{\partial\mu}{\partial t}=\dfrac{1}{L_\mu(1-k^2)}\left(\dfrac{\partial^2\mu}{\partial z^2}+\dfrac{M}{L_c}\dfrac{\partial^2c}{\partial z^2}\right).
\end{equation}
Since the circuits of $\mu$ and $c$ are structurally identical, we may switch their roles to find 
\begin{equation}
\dfrac{1}{R_c}\dfrac{\partial c}{\partial t}=\dfrac{1}{L_c(1-k^2)}\left(\dfrac{M}{L_\mu}\dfrac{\partial^2\mu}{\partial z^2}+\dfrac{\partial^2c}{\partial z^2}\right). 
\end{equation}
\label{eq:tl_eqns}
\end{subequations}
\eqref{eq:tl_eqns} are of the same form as the averaged Poisson-Nernst-Planck \eqref{eq:avg_pnp} for the resistances and inductances given in terms of the electrolyte properties in Table \ref{tab:circuit_properties}. Note that when $\beta=0$, $M=0$, hence the two circuits decouple and the generalized de Levie circuit is recovered.

\begin{table}[h!]
\centering
\caption{Elements of magnetically coupled LR circuits for EDL charging of asymmetric electrolytes in a slender pore. The circuit is provided in Fig.~\ref{fig:circuits}b.}
\begin{tabular}{ccccc}
\hline\\[-10pt]
$R_\mu$               & $L_\mu$                            & $R_c$                   & $L_c$                                            & $M$                        \\ \\[-10pt]\hline\\[-7pt]
$\mathcal{D}(\kappa)$ & $\dfrac{1-\beta\gamma}{1-\beta^2}$ & $\dfrac{1}{1-\gamma^2}$ & $\dfrac{1+\beta\gamma}{(1-\beta^2)(1-\gamma^2)}$ & $\dfrac{\beta}{1-\beta^2}$ \\ [10pt]\hline
\end{tabular}
\label{tab:circuit_properties}
\end{table}

Summarizing, the magnetically coupled circuits we have derived describe the slow-fast charging dynamics of asymmetric electrolytes in confinement due to differing ionic mobilities. The impedance of each element is multiplied by $i\omega$ compared to de Levie's circuits for charge and salt. This allows for introducing magnetic couplings between the neighboring inductors of different circuits, representing salt accumulation by charge gradients and charge accumulation by salt gradients. Importantly, other circuit elements used in electrochemistry can be added to the magnetically coupled circuits by multiplying their impedances by $i\omega$ and determining the impedance of the resulting electrochemical system from the time derivative of the current in the circuit, as illustrated in the simple example. However, it should be noted that this circuit element transformation we propose would map the impedance of inductors to $-\omega^2L$, which doesn't correspond to a traditional circuit element. This shouldn't be a hurdle to the application of the model to ion transport since inductors are infrequent in electrochemical circuits in the time domain, generally ascribed to impedance from cell cables and otherwise restricted to the chemical formation of intermediate species \cite{lazanas2023electrochemical}. Finally, it should be noted that the inductors in the time-differentiated circuit we propose do not imply electric fields produced by time-varying magnetic fields in the electrolyte; rather, they are a tool to represent the impedance response of the confined electrolyte through a circuit with modified elements. In the physical system, it comes from coupled potential gradients and currents (not their time derivatives) of salt density and electrochemical potential of charge.

\section{Impedance Response}

We examine the impedance response of EDL charging due to the coupled charge and salt dynamics expressed in the proposed circuit directly through \eqref{eq:avg_pnp}. It should be noted that the linearization of the PNP equations formally consists of collecting first-order terms in the asymptotic expansion of these equations for a small potential magnitude, i.e., for $\epsilon\ll 1$. Therefore, in the solution of the equations, we will rescale all independent variables using the amplitude of the applied electric potential instead of the thermal voltage. The only change this rescaling effects on the problem formulation is the expression of the electrode potential, which now has unit magnitude, $\phi_D(t)=\sin(\omega t)$. We look for steady-state solutions of the form $\mu(z,t)=\textrm{Im}(\hat{\mu}(z)\exp(i\omega t))$ for all dependent variables. Thus, \eqref{eq:avg_pnp} sets the ordinary differential equations governing the hat variables (spatial dependences),
\begin{subequations}
    \begin{equation}
        \dfrac{i\omega}{\mathcal{D}}\hat{\mu}=(1+\beta\gamma)\dfrac{d^2\hat{\mu}}{dz^2}+\beta(1-\gamma^2)\dfrac{d^2\hat{c}}{dz^2}
        \label{eq:mu_hat}
    \end{equation}
    and
    \begin{equation}
        i\omega\hat{c}=\beta\dfrac{d^2\hat{\mu}}{dz^2}+(1-\beta\gamma)\dfrac{d^2\hat{c}}{dz^2}.
        \label{eq:c_hat}
    \end{equation}
    \label{eq:hat}
\end{subequations}
Applying the operator $\beta(1-\gamma^2)d^2/dz^2$ to \eqref{eq:c_hat} and substituting the resulting derivatives of salt by their corresponding expressions in \eqref{eq:mu_hat},
we find a single fourth-order ODE for the spatial dependence of the excess electrochemical potential of charge,
\begin{subequations}    
\begin{equation}
    \dfrac{d^4\hat{\mu}}{dz^4}-\dfrac{i\omega}{1-\beta^2}\left(1+\beta\gamma+\dfrac{1-\beta\gamma}{\mathcal{D}}\right)\dfrac{d^2\hat{\mu}}{dz^2}+\dfrac{i^2\omega^2\hat{\mu}}{(1-\beta^2)\mathcal{D}}=0.
    \label{eq:mu}
\end{equation}
For simplicity and physical intuition, we discuss the case of a negligible SDL resistance, where the measured impedance is entirely due to the pore. In the Appendix, we will add the SDL resistance to the analysis. We apply the blocking-end boundary conditions $d\hat\mu/dz(z=1)=d\hat c/dz(z=1)=0$, written in terms of electrochemical potential of charge using \eqref{eq:hat} as
\begin{equation}
    \dfrac{d\hat{\mu}}{dz}\bigg|_{z=1}=\dfrac{d^3\hat{\mu}}{dz^3}\bigg|_{z=1}=0,
    \label{eq:blocking_mu}
\end{equation}
and the reservoir boundary conditions for charge and salt, namely
\begin{equation}
    \hat{\mu}(z=0)=-1,
\end{equation}
which is the dimensionless excess of the electric potential of the reservoir relative to the electrode, and
\begin{equation}
    \left(\dfrac{d^2\hat{\mu}}{dz^2}-\dfrac{1-\beta\gamma}{1-\beta^2}\dfrac{i\omega}{\mathcal{D}}\hat{\mu}\right)\bigg|_{z=0}=0
\end{equation}
from the vanishing excess salt of the reservoir, $\hat{c}(z=0)=0$, written in terms of the electrochemical potential of charge using \eqref{eq:hat}.
\label{eq:bvp_mu}
\end{subequations}
The general solution of \eqref{eq:mu} with the blocking-end boundary conditions is
\begin{subequations} 
\begin{equation}
    \hat{\mu}=C_+\cosh(m_+(z-1))+C_-\cosh(m_-(z-1)),
\end{equation}
where $C_\pm$ are constants of integration and
\begin{equation}
    m_\pm=\sqrt{\dfrac{i\omega[(1+\beta\gamma)\mathcal{D}+1-\beta\gamma]}{2(1-\beta^2)\mathcal{D}}}\left\{1\pm\left(1-\dfrac{4(1-\beta^2)\mathcal{D}}{[(1+\beta\gamma)\mathcal{D}+1-\beta\gamma]^2}\right)^{1/2}\right\}^{1/2}
\end{equation}
are the roots of the biquadratic characteristic equation of \eqref{eq:mu},
\begin{equation}
    m^4-\dfrac{i\omega}{1-\beta^2}\left(1+\beta\gamma+\dfrac{1-\beta\gamma}{\mathcal{D}}\right)m^2+\dfrac{i^2\omega^2}{(1-\beta^2)\mathcal{D}}=0.
\end{equation}
$m_\pm$ can be interpreted as complex-valued penetration wavenumbers, as in solutions of the RC transmission-line model \cite{orazem2017electrochemical}. It should be noted that the penetration wavenumbers are proportional to $\sqrt{i\omega}$, which ensues a 45$^\circ$-line asymptote for high frequencies. $m_\pm$ are also related to the eigenvalues $\lambda_\pm$ determined in Ref. \cite{henrique2022impact}, related to charging frequencies. Applying the reservoir boundary conditions, we find
\begin{equation}
     C_+=\dfrac{\mathcal{C}}{(\mathcal{B}-\mathcal{C})\cosh(m_+)},\quad
     C_-=-\dfrac{\mathcal{B}}{(\mathcal{B}-\mathcal{C})\cosh(m_-)},
\end{equation}
where
\begin{equation}
    \mathcal{B}=m_+^2-\dfrac{i\omega}{\mathcal{D}}\dfrac{1-\beta\gamma}{1-\beta^2}\quad\mathrm{and}\quad
    \mathcal{C}=m_-^2-\dfrac{i\omega}{\mathcal{D}}\dfrac{1-\beta\gamma}{1-\beta^2}.
\end{equation}
\label{eq:constants}
\end{subequations}
\par{}The dimensionless impedance is calculated as the inverse of the charge flux at the pore inlet \cite{pedersen2023equivalent}, which can be inferred from the right-hand side of \eqref{eq:mu_hat} and written in terms of derivatives of the electrochemical potential of charge using \eqref{eq:c_hat}. The impedance can then be evaluated as
\begin{subequations}
\begin{align}  \mathcal{Z}&=1\bigg/\left[\left(1+\beta\gamma+\dfrac{1-\beta\gamma}{\mathcal{D}}\right)\dfrac{d\hat{\mu}}{dz}+\dfrac{i(1-\beta^2)}{\omega}\dfrac{d^3\hat{\mu}}{dz^3}\right]\bigg|_{z=0}\nonumber\\
&=\left[\dfrac{m_+}{A_+(\kappa,\beta,\gamma)\coth(m_+)}+\dfrac{m_-}{A_-(\kappa,\beta,\gamma)\coth(m_-)}\right]^{-1}.
\label{eq:Z}
\end{align}
Here, $A_\pm$ are constants that depend on the electrolyte parameters and relative pore size, given by
\begin{equation}
    A_+=\dfrac{m_+^2-m_-^2}{\mathcal{B}}\left[1+\beta\gamma+\dfrac{1-\beta\gamma}{\mathcal{D}}+\dfrac{i(1-\beta^2)}{\omega}m_-^2\right]^{-1}
\end{equation}
and
\begin{equation}
    A_-=-\dfrac{m_+^2-m_-^2}{\mathcal{C}}\left[1+\beta\gamma+\dfrac{1-\beta\gamma}{\mathcal{D}}+\dfrac{i(1-\beta^2)}{\omega}m_+^2\right]^{-1}.
\end{equation}
\label{eq:Z_pore}
\end{subequations}
\par{} The presence of two such wavenumbers, in contrast to the single penetration depth of de Levie's model, reflects the two transport timescales owing to the coupled dynamics of charge and salt. \eqref{eq:Z} shows that two parallel Warburg open impedances, corresponding to the two modes of transport, can be combined to furnish the net impedance of the electrolyte in the pore when the bulk electrolyte resistance is negligible. When the bulk electrolyte resistance is not negligible, additional Warburg elements appear; see Appendix.

\begin{figure}[h!]
    \centering
    \includegraphics[width=.6\textwidth]{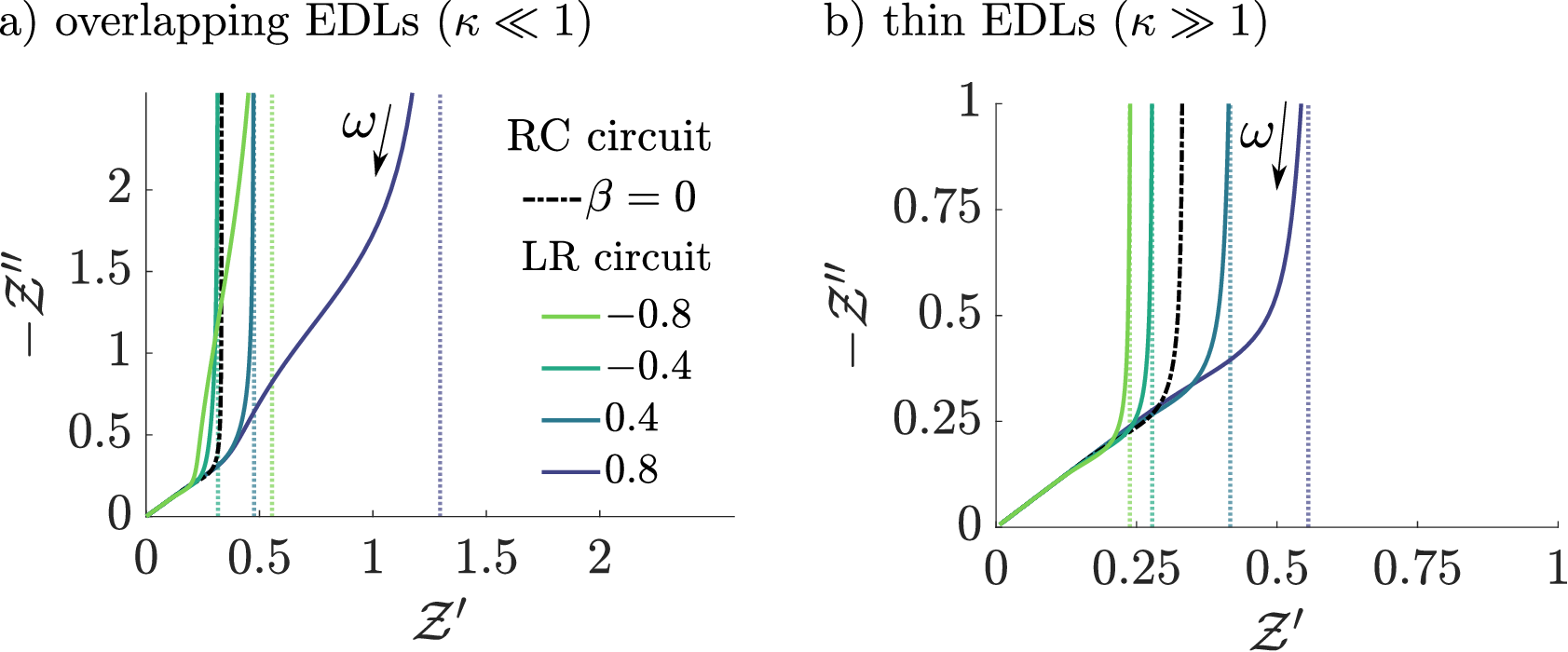}
    \caption{\textbf{Impact of diffusivity asymmetry on the Nyquist plots (\eqref{eq:Z}) with $\gamma=0.5$} for a) overlapping and b) thin EDLs. Low-frequency capacitive asymptotes are represented by dotted lines (Eq. \ref{eq:Z_pore} for $\omega\ll 1$). Two turns are observed in the overlapping double-layer regime with electrolyte asymmetry in contrast to a single turn in the thin double-layer regime.}
    \label{fig:nyquist}
\end{figure}

The impact of electrolyte asymmetry on the impedance is illustrated 
through the Nyquist plots in Fig. \ref{fig:nyquist}. As described earlier, the results differ from strongly coupled circuits for overlapping EDLs to weakly coupled ones for thin EDLs. Fig.~\ref{fig:nyquist}a shows the plots for overlapping EDLs. The high-frequency response retains the character of a semi-infinite Warburg element, with a 45$^\circ$-line asymptote for all parameters shown. From moderate to low frequencies, a high diffusivity asymmetry introduces a qualitatively different approach to the capacitive low-frequency vertical asymptote. For a high diffusivity of the ion with lower valence ($\beta=-0.8$, $\gamma=0.5$), the intermediate frequency regime resembles a constant-phase element (CPE). The low-frequency resistance limits relate to a weighted diffusivity where the weights are the valences of the corresponding ions. This contrasts with thin EDLs, where the resistance is set by the ion diffusivities weighted by valences of the opposing ions, i.e., the ambipolar diffusivity. This difference stems from the breakdown of electroneutrality in the narrow pores. Near equilibrium, the salt density is undisturbed, but the uniformity of electrochemical potentials requires the charge distribution to match the slowly evolving unscreened electric potentials. Diffusive transport in this regime is controlled not by variations of salt, but of charge, in opposition to the electroneutrality of thin EDLs. This distinction is significant for the inference of electrolyte properties from Nyquist plots in EIS. Thus, our model suggests that features of the impedance response may have a previously unrecognized dependence on pore size. In Fig. \ref{fig:nyquist}b, we observe that the thin EDLs impedance plots have the shape of a single Warburg open element, with the well-known low-frequency resistance limit corresponding to the ambipolar diffusivity. 

\section{Conclusion}

Here, we have developed a circuit interpretation of EDL charging of asymmetric electrolytes in a slender pore. This effect required a modified circuit representation for overlapping double layers since unequal ion mobilities in a charged electrolyte produce coupled dynamics of charge density and total ion concentration (salt density). To provide a circuit representation of this phenomenon, we interpreted the mechanistic ion transport equations of Ref. \cite{henrique2022impact} as circuit equations for two ``voltages'': the electrochemical potential of charge and the salt density. In this language, either voltage drop produces a current in both circuits. To describe this coupling, we interpreted the governing equations as coupled de Levie circuits for the electrochemical potential of charge and salt density with element impedances multiplied by $i\omega$. That is, with capacitors replaced by resistors, resistors replaced by inductors, and mutual inductances added between the corresponding inductors of different circuits. We discussed through an example how the modified coupled circuits are consistent with mapping the physical currents of charge and salt to the time derivatives of these currents in the circuits. This counterintuitive association is required to capture the effects of salt gradients in charge currents. Therefore, the inference of the impedance response from a circuit with inductors does not imply the significance of time-varying magnetic fields; it is rather a tool to represent slow-fast coupled charge and salt dynamics through mutual inductances.

As in other transmission line models, important limitations of the model arise from its basis on linearized Poisson-Nernst-Planck equations. Namely, it does not account for large electrode potentials \cite{bazant2004diffuse}, finite ion-size effects \cite{kilic2007steric1,gupta2018electrical}, ion-ion correlations \cite{storey2012effects}, or highly concentrated electrolytes \cite{lee2015dynamics}. However, the circuit can be extended to include other electrochemical effects commonly considered in TL fits of EIS measurements. To this end, one can either construct a circuit with the impedance of all elements multiplied by $i\omega$ and infer the electrolyte impedance from the time derivative of the current through the circuit, or use parallel Warburg elements to represent the pore impedance. 
We believe this representation will help model impedance responses of porous electrodes with coupled exponential relaxations. The work here considers binary electrolytes, but multicomponent electrolyte transport can introduce more timescales and coupled equations, which could introduce more impedance oscillations.  Furthermore, the results suggest that ionic transport in electrochemical processes such as  electroconvection~\cite{davidson2016dynamical,ratschow2024convection}, surface-enhanced diffusion in nanopores~\cite{christensen2023locally}, and bipolar membranes~\cite{zhang2024modulation} should investigate charge and salt coupling due to diffusivity asymmetry. The parallel Warburg frequency response characterized in this paper can also be leveraged in the manipulation of EDLs for producing nanofluidic electrokinetic flows controlled by gated potentials \cite{ratschow2022resonant} or electrochemical reactions in patchy pore walls \cite{shrestha2024self}. Excitingly, the work also suggests a possible link between electrolyte transport and magnetic circuits, for instance in chemical inductors where a fast and a slow mode is required~\cite{bisquert2022chemical} and for systems with multiple salt gradients~\cite{warren2024salt}.


\section*{Acknowledgment}
A.G. thanks the NSF (CBET-2238412) CAREER award for financial support. F.H. thanks the Ryland Graduate Family Fellowship for financial assistance. 

\appendix

\setcounter{figure}{0}
\renewcommand\thefigure{A\arabic{figure}}
\section{Effect of SDL Resistance}

Here, we extend the calculation of the main text to include the resistance of the SDL. The point of this exercise is to show that the impedance response of the electrolyte may be more convoluted than a parallel combination of Warburg open elements. This response can still be predicted by the transmission-line equations and an equivalent circuit representation, which must now include a frequency-shifted transmission line for the SDL. However, it is harder to reconstruct by guessing a combination of Warburg elements.

The same formulation of the pore region applies, but now we need to append the SDL to the domain by an extension of the axial coordinate, i.e., $-\ell_s<z<0$. In the SDL, we assume that $\rho=0$, such that the transport equations effectively correspond to the limit $\kappa\to \infty$ of \eqref{eq:avg_pnp}. In terms of the excess electrochemical potential of charge, the spatial part of the periodic response is determined from the substitution of \eqref{eq:mu_hat} into the second derivative of \eqref{eq:c_hat} in the electroneutral limit, with $\mathcal{D}\to\infty$, as
\begin{subequations}
\begin{equation}
    i\omega\dfrac{d^2\hat{\mu}}{dz^2}=D_a\dfrac{d^4\hat{\mu}}{dz^4},
    \label{eq:epoc_sdl}
\end{equation}
where $D_a=(1-\beta^2)/(1+\beta\gamma)$ is the dimensionless ambipolar diffusivity. The boundary conditions of the electroneutral reservoir with undisturbed salt concentration now apply 
to the start of SDL,
\begin{equation}
    \hat{\mu}(z=-\ell_s)=-1        
\end{equation}
and
\begin{equation}
    \dfrac{d^2\hat{\mu}}{dz^2}\bigg|_{z=-\ell_s}=0.
\end{equation}
To complete the set of boundary conditions of both fourth-order equations, four additional matching conditions are required. They establish the continuity of the electrochemical potential of charge and of salt density, as well as the conservation of the currents of charge and salt. The continuity conditions written in terms of electrochemical potentials are using similar manipulations as before, as
\begin{equation}
    \hat{\mu}(z=0^-)=\hat{\mu}(z=0^+)
    \label{eq:mu_continuous}
\end{equation}
and
\begin{equation}
     \dfrac{d^2\hat{\mu}}{dz^2}\bigg|_{z=0^-}=\left(\dfrac{d^2\hat{\mu}}{dz^2}-\dfrac{i\omega}{\mathcal{D}}\dfrac{1-\beta\gamma}{1-\beta^2}\hat{\mu}\right)\bigg|_{z=0^+}.
\end{equation}
Similarly, bearing in mind the difference in the cross-sectional areas of the SDL $A_s$ and the pore inlets $A$, which we approximate by the porosity of the medium, $\vartheta\approx A/A_s$, the conservation equations of charge and salt currents are given by
\begin{equation}
    \dfrac{1}{\vartheta}\dfrac{d\hat{\mu}}{dz}\bigg|_{z=0^-}=\dfrac{d\hat{\mu}}{dz}\bigg|_{z=0^+}
\end{equation}
and
\begin{equation}
    \dfrac{1}{\vartheta}\dfrac{d^3\hat{\mu}}{dz^3}\bigg|_{z=0^-}=\left(\dfrac{d^3\hat{\mu}}{dz^3}-\dfrac{i\omega}{\mathcal{D}}\dfrac{1-\beta\gamma}{1-\beta^2}\dfrac{d\hat{\mu}}{dz}\right)\bigg|_{z=0^+}.
    \label{eq:salt_current}
\end{equation}
\label{eq:mu_SDL}
\end{subequations}
In summary, we must solve the boundary-value problem given by \eqref{eq:mu} and \eqref{eq:blocking_mu} and \eqref{eq:mu_SDL}. General solutions to this governing equations that satisfy the reservoir and blocking-end boundary conditions are of the form
\begin{equation}
    \hat{\mu}=\begin{cases}
        C_1\sinh(m_s(z+\ell_s))+C_2(z+\ell_s)-1,\quad -\ell_s\le z\le 0,\\
        C_3\cosh(m_+(z-1))+C_4\cosh(m_-(z-1)),\quad 0\le z\le 1,
    \end{cases}
    \label{eq:varphi_with_sdl}
\end{equation}
where $m_s=\sqrt{i\omega/D_a}$. To determine the impedance, we need to calculate the inverse of the current at the mouth of the pore, on either the SDL or the pore side of this microscopically diffuse interface. We choose the former, such that the dimensionless impedance based on the area of the pore is determined from the charge flux inferred from \eqref{eq:hat} as
\begin{subequations}
\begin{equation}
    \mathcal{Z}=\vartheta\left.\left[(1+\beta\gamma)\dfrac{d\hat{\mu}}{dz}+\dfrac{i}{\omega}(1-\beta^2)\dfrac{d^3\hat{\mu}}{dz^3}\right]^{-1}\right|_{z=0^-}.
\end{equation}
In terms of the solution of the electrochemical potential of charge in the SDL, it is given by
\begin{equation}
    \mathcal{Z}=\dfrac{\vartheta}{1+\beta\gamma}\left\{\left[1+\dfrac{i(1-\beta^2)m_s^2}{(1+\beta\gamma)\omega}\right]C_1m_s\cosh(m_s\ell_s)+C_2\right\}^{-1}=\dfrac{\vartheta}{(1+\beta\gamma)C_2}.
    \label{eq:impedance_sdl}
\end{equation}
\end{subequations}

\begin{figure}[h!]
    \centering
    \includegraphics[width=.3\textwidth]{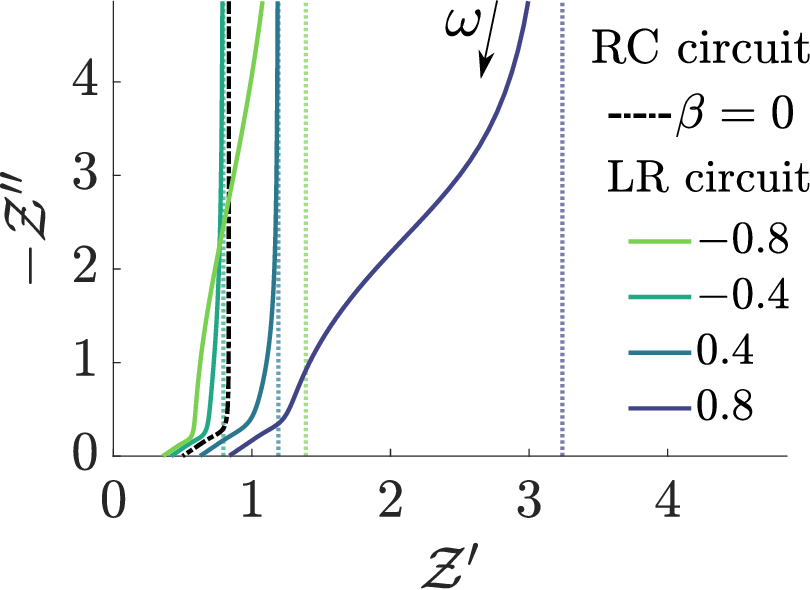}
    \caption{\textbf{Impact of a static diffusion layer on the Nyquist plots (\eqref{eq:impedance_sdl} and \eqref{eq:coeffs}) with $\gamma=0.5$, $\ell_s=0.5$, and $\vartheta=0.5$.} for overlapping EDLs. Only two turns are observed in the overlapping double-layer regime with electrolyte asymmetry, despite the SDL resistance.}
    \label{fig:nyquist_sdl}
\end{figure}

The interfacial boundary conditions that remain, \eqref{eq:mu_continuous}--\eqref{eq:salt_current}, are written in terms of the constants of integration as the following linear system of equations
\begin{equation}
    \begin{cases}
        \sinh(m_s\ell_s)C_1+\ell_sC_2-\cosh(m_+)C_3-\cosh(m_-)C_4=1,\\[10pt]
        \dfrac{m_s}{\vartheta}\cosh(m_s\ell_s)C_1+\dfrac{1}{\vartheta}C_2+m_+\sinh(m_+)C_3+m_-\sinh(m_-)C_4=0,\\[10pt]
        m_s^2\sinh(m_s\ell_s)C_1-\left(m_+^2-\dfrac{1-\beta\gamma}{1-\beta^2}\dfrac{i\omega}{\mathcal{D}}\right)\cosh(m_+)C_3-\left(m_-^2-\dfrac{1-\beta\gamma}{1-\beta^2}\dfrac{i\omega}{\mathcal{D}}\right)\cosh(m_-)C_4=0,\\[10pt]
        \dfrac{m_s^3}{\vartheta}\cosh(m_s\ell_s)C_1+\left(m_+^2-\dfrac{1-\beta\gamma}{1-\beta^2}\dfrac{i\omega}{\mathcal{D}}\right)m_+\sinh(m_+)C_3+\left(m_-^2-\dfrac{1-\beta\gamma}{1-\beta^2}\dfrac{i\omega}{\mathcal{D}}\right)m_-\sinh(m_-)C_4=0.\\
    \end{cases}
\end{equation}
Solving by Kramer's rule, we obtain the required coefficient,
\begin{subequations}
\begin{equation}
    C_2=\dfrac{\det(M_2)}{\det(M)},
    \label{eq:C2}
\end{equation}
where
\begin{equation}
    M=\begin{bmatrix}
        \sinh(m_s\ell_s)& \ell_s & -\cosh(m_+)& -\cosh(m_-)\\[10pt]
        \dfrac{m_s}{\vartheta}\cosh(m_s\ell_s)&\dfrac{1}{\vartheta}& m_+\sinh(m_+) & m_-\sinh(m_-)\\[10pt]
        m_s^2\sinh(m_s\ell_s) & 0 & -\mathcal{B}\cosh(m_+) & -\mathcal{C}\cosh(m_-)\\[10pt]
        \dfrac{m_s^3}{\vartheta}\cosh(m_s\ell_s) & 0 & \mathcal{B}m_+\sinh(m_+) & \mathcal{C}m_-\sinh(m_-)
    \end{bmatrix}
\end{equation}
is the matrix of coefficients of the linear system, and
$M_i$ is the matrix obtained by the substituting the $i$-th column of $M$ by the right-hand side vector $\begin{bmatrix}
1 & 0 & 0 & 0\end{bmatrix}^T$ of the system. Defining the shorthand $\mathcal{A}=m_s^2$ and calculating the determinants, we find
\begin{align}
    \det(M)=\,\mathcal{A}\sinh(m_s\ell_s)\bigg\{&\ell_sm_+m_-\sinh(m_+)\sinh(m_-)(\mathcal{C}-\mathcal{B})\nonumber\\&+\dfrac{1}{\vartheta}\left[m_-\cosh(m_+)\sinh(m_-)\mathcal{C}-m_+\cosh(m_-)\sinh(m_+)\mathcal{B}\right]\bigg\}\nonumber\\
    +\dfrac{\mathcal{B}}{\vartheta}\cosh(m_+)\bigg\{&\ell_sm_sm_-\cosh(m_s\ell_s)\sinh(m_-)(\mathcal{C}-\mathcal{A})\nonumber\\&-\dfrac{m_s}{\vartheta}\cosh(m_-)\cosh(m_s\ell_s)\mathcal{A}-m_-\sinh(m_s\ell_s)\sinh(m_-)\mathcal{C}\bigg\}\nonumber\\
    -\dfrac{\mathcal{C}}{\vartheta}\cosh(m_-)\bigg\{&\ell_sm_sm_+\cosh(m_s\ell_s)\sinh(m_+)(\mathcal{B}-\mathcal{A})\nonumber\\&-\dfrac{m_s}{\vartheta}\cosh(m_+)\cosh(m_s\ell_s)\mathcal{A}-m_+\sinh(m_s\ell_s)\sinh(m_+)\mathcal{B}\bigg\},
\end{align}
and
\begin{align}
\det(M_2)=\mathcal{A}m_+m_-\sinh(m_s\ell_s)\sinh(m_+)\sinh(m_-)(\mathcal{C}-\mathcal{B})\nonumber\\
+\dfrac{\mathcal{B}}{\vartheta}m_sm_-\cosh(m_s\ell_s)\cosh(m_+)\sinh(m_-)(\mathcal{C}-\mathcal{A})\nonumber\\
-\dfrac{\mathcal{C}}{\vartheta}m_sm_+\cosh(m_s\ell_s)\sinh(m_+)\cosh(m_-)(\mathcal{B}-\mathcal{A}).
\end{align}
\label{eq:coeffs}
\end{subequations}
Since the division in \eqref{eq:C2} brings about a counterintuitive combination of hyperbolic tangents of the penetration wavenumbers $m_\pm$ and $m_s\ell_s$, we conclude that the impedance given by \eqref{eq:impedance_sdl} and \eqref{eq:coeffs} is a more convoluted combination of Warburg elements. Mathematically, introducing a static diffusion layer does not amount to simply adding a resistance to the impedance. However, the Nyquist plots illustrated in Fig. \ref{fig:nyquist_sdl} only show two turns, with a similar qualitative response to the negligible SDL resistance.

\end{document}